\documentclass[twocolumn,showpacs,preprintnumbers,amsmath,amssymb]{revtex4}
\usepackage{graphics}
\begin{document}
\title{Absence of Quantum Metallic Behavior in Disordered Granular Superconductors}

\author{ Ryusuke Ikeda }

\affiliation{%
Department of Physics, Kyoto University, Kyoto 606-8502, Japan 
}%

\date{\today}

\begin{abstract}
We examine the idea, postulated by Phillips et al., that a finite 
resistivity in $T \to 0$ limit in disordered {\it granular} superconducting (SC) films is explained as a consequence of the absence of phase stiffness in a phase glass (PG) peculiar to granular systems. It is found that, in spite of the absence of static phase stiffness, a coupling between the nonzero PG order and the ordinary SC fluctuation makes the conductivity divergent. However, an actual drop of resistance is argued to occur due to another SC glass ordering, induced by the precursory PG {\it fluctuation}, corresponding to the vortex-glass transition in a nonzero magnetic field. 

\end{abstract}


\maketitle
%
%

Study of the phase diagram in disordered granular superconductors at low temperature ($T$) is important in relation to understanding the superconductor-insulator (S-I) transition behaviors. Recently, Dalidovich and Phillips \cite{Phillips,Phillips2} have argued that, in a two dimensional (2D) phase glass (PG) expected to be induced in low $T$ limit by a disordered granular structure, the resistance is finite. Observations of a quantum metallic behavior \cite{Goldman,Goldman2,Kapitulnik} do not seem to be directly explained within available theories \cite{RI,else,RI2} for {\it homogeneously} disordered superconducting (SC) materials. Further, it is found that their analysis can be trivially extended to the 3D case and the case with nonzero magnetic field ($H \neq 0$) and leads to a similar quantum metallic behavior. Since resistive data showing a metallic resistance curves flattening upon cooling are available in quasi 2D systems \cite{3D} in $H \neq 0$, their proposal, if correct, might become a correct description of the corresponding phenomena. On the other hand, it has been argued \cite{GaL} that, in contrast to the homogeneously-disordered case \cite{RI2}, the resistivity drop at low enough (but finite) temperatures in disordered superconductors consisting of clustered SC islands (i.e., SC grains) should occur even in a wide field range {\it above} an averaged (mean field) upper critical 
field ${\overline H}_{c2}(0)$ . Therefore, effects of disordered granular structure on quantum resistive behaviors are not theoretically understood well at present. 

To examine this problem, we use essentially the same model as in Ref.\cite{Phillips} expressing 
a random Josephson junction array with an on-site charging energy. 
\begin{equation}
{\cal S}= {\cal S}_0 - \frac{1}{2} \int^\beta_0 d\tau \sum_{<i,j>} \biggl({\tilde J}_{ij}(\tau) S_{+i}(\tau) S_{-j}(\tau) + c.c. \biggr),
\end{equation}
where the pair $<i$, $j>$ denotes a nerest-neighbor pair, and 
\begin{eqnarray}
{\cal S}_0 &=& \int^\beta_0 d\tau \sum_i \frac{1}{2 \alpha} \biggl(\frac{\partial \theta_i(\tau)}{\partial \tau} \biggr)^2, \\ \nonumber
\cr {\tilde J}_{ij}(\tau) &=& {\tilde J}_{ji}^*(\tau) = J_{ij} \, \exp{[ \, {\rm i} ({\bf A}_{ij}^{\rm ext} + \delta {\bf A}_{ij}(\tau)) ]}, \\ \nonumber
\cr S_{\pm i}(\tau) &=& \exp(\pm{\rm i} \, \theta_i(\tau)). 
\end{eqnarray}

The gauge field consists of the static component ${\bf A}_{ij}^{\rm ext}$ for an applied uniform magnetic field ${\bf H}$ and the dynamical fluctuation $\delta {\bf A}(\tau)$ introduced for deriving the linear responses. The quenched disorder in the system is incorporated in a randomness of $J_{ij}$ with nonzero mean $J_0$ (${\overline {J_{ij}}}=J_0 > 0$) for any $<i$, $j>$ and with a Gaussian distribution. Only for a formal justification of the mean-field approach on the PG ordering, the Gaussian distribution on $J_{ij} - J_0$ will be replaced by an infinite-ranged one. Then, after replicating the action and introducing PG and SC order parameter fields, $q^{ab}(\tau_1,\tau_2)=(q^{ba}(\tau_2, \tau_1))^*$ and $\psi_i^a(\tau)$, the random-averaged free energy is given by ${\rm lim}_{n \to +0} ({\overline Z^n} - 1)/n$, where 
\begin{equation}
\frac{\overline {Z^n}}{\overline {Z^n_0}} = \int {\cal D}\psi {\cal D}\psi^* {\cal D} q \exp(-{\cal F}_{\rm eff}(\psi, q)), 
\end{equation}
${\overline {Z^n_0}}$ is the replicated partition function for ${\cal S}_0$, 
\begin{eqnarray}
{\cal F}_{\rm eff}(\psi, q) = \frac{N}{2 J^2} \! \! \int d\tau_1 \! \! \int &d\tau_2& \sum_{a,b} |q^{(ab)}(\tau_1, \tau_2)|^2  \\ \nonumber
\cr \times \biggl( 1 + \frac{(\delta A(\tau_1) - \delta A(\tau_2))^2}{2} \biggr) &+& \frac{1}{4d} \sum_i \int d\tau \sum_a |\psi_i^{(a)}|^2 \\ \nonumber
\cr - {\rm ln} \biggl[ \biggl< T_\tau \exp \biggl(\frac{\sqrt{J_0}}{2} \int &d\tau& \sum_a \sum_i S_{+_i}^{(a)}(\tau) \\ \nonumber 
\cr \times \biggl(\frac{{\bf D}_i \cdot {\bf D}_i^*}{2d} + &1& \biggr)^{1/2} 
(\psi_i^{(a)}(\tau))^* \\ \nonumber
\cr + \frac{1}{2} \int d\tau_1 d\tau_2 \sum_{a,b} q^{(ba)}(\tau_2, \tau_1&)& \sum_i S_{+i}^{(a)}(\tau_1) S_{-i}^{(b)}(\tau_2) 
+ {\rm c.c.} \biggr>_0 \biggr],
\label{repact}
\end{eqnarray}
and the bracket $\langle \,\,\, \rangle_0$ denotes the ensemble average by the charging energy part ${\cal S}_0$. 
In eq.(4), $a$ and $b$ are replica indices, $d$ is the spatial dimension, ${\bf D}_i \cdot {\bf D}_i^*$ denotes the Laplacian on the gauge-invariant gradients defined on the cubic lattice (for $d=3$), and the factor $N$ (the system size) in the first and second terms arises due to the replacement into the infinite-ranged model. The gauge fluctuation $\delta {\bf A} = \delta A {\hat x}$ applied externally is assumed to be small, and its site dependence was neglected because spatially uniform linear responses are considered below. Performing the cumulant expansions with respect to $q^{(ab)}$ and $\psi^{(a)}$, ${\cal F}_{\rm eff}$ is obtained in a form of a Landau free energy functional. Further, as in Ref.\cite{Read,Dalid}, rewriting $q^{(ab)}(\tau_1, \tau_2)$ as $Q^{(ab)}(\tau_1, \tau_2) - C \delta_{a,b} \delta(\tau_1-\tau_2)$ in order to delete the term $\int d\tau_1 \int d\tau_2 |Q^{(ab)}(\tau_1, \tau_2)|^2$, we finally obtain the following effective action

\begin{eqnarray}
\frac{t {\cal F}_{\rm eff}(\psi, Q)}{N} &=& \kappa^{-1} \int d\tau \sum_a \biggl( \frac{\partial^2}{\partial \tau_1 \, \partial \tau_2} + r \biggr) Q^{(aa)}(\tau_1, \tau_2) \biggr|_{\tau_1=\tau_2} \\ \nonumber 
\cr - \frac{\kappa}{3} \int d\tau_1 d\tau_2 &d\tau_3& \sum_{a,b,c} Q^{(ab)}(\tau_1,\tau_2) Q^{(bc)}(\tau_2,\tau_3) Q^{(ca)}(\tau_3, \tau_1) \\ \nonumber
\cr + \frac{u}{2} \int d&\tau& \sum_a (Q^{(aa)}(\tau,\tau))^2 
+ \frac{t {\cal F}_\psi}{N} + \frac{t {\cal F}_A}{N}, 
\end{eqnarray}
where 
\begin{equation}
\frac{t {\cal F}_A}{N} = \frac{t}{4 J^2} \sum_{a,b} \int d\tau_1 d\tau_2 (\delta A(\tau_1)- \delta A(\tau_2))^2 |Q^{(ab)}(\tau_1, \tau_2)|^2,
\end{equation}

\begin{eqnarray}
t {\cal F}_\psi &=& \sum_i \biggl[ \int d\tau \sum_a \biggl( r_\psi |\psi_i^{(a)}(\tau)|^2 \\ \nonumber 
\cr &+& c_\psi \biggl|\frac{\partial \psi_i^{(a)}}{\partial \tau} \biggr|^2 + c_g |D_i \psi_i^{(a)}|^2 \biggr ) \\ \nonumber
\cr  - w_\psi \sum_{a,b} \int d&\tau_1& \int d\tau_2 (\psi_i^{(a)}(\tau_1))^* Q^{(ab)}(\tau_1, \tau_2) \psi_i^{(b)}(\tau_2) \biggr].
\end{eqnarray}
For a moment, a $|\psi|^4$ term with positive coefficient will be neglected. The coefficients in eqs.(5) and (6) can be calculated from correlation functions on $S_\pm(\tau)$ on the basis of the local action ${\cal S}_0$, and, except $r$ and $r_\psi$, all of them are always positive . For instance, the coefficient $w_\psi$ is postive as far as $J_0 > 0$. The above expression of the action is of the same form as in Ref.\cite{Dalid}. 

Now, let us examine the mean field solution on $Q^{ab}$ by taking account of the presence of SC ($\psi$-) fluctuation. In $H \neq 0$ case, this treatment may be appropriate {\it above} ${\overline H}_{c2}(0)$ and in low enough temperatures \cite{Lubensky}. Note that there is no reason why the SC fluctuation is negligible in the mean field analysis of the PG order because the $\psi$-field in eq.(7) couples to the $Q$-field in a bilinear form $(\psi^{(a)})^* \psi^{(b)}$. As shown below, including SC fluctuations is essential to obtaining a correct result of conductivity. Following Ref.\cite{Read,Dalid}, the mean field ansatz $Q^{(ab)}(\tau_1,\tau_2) =  q (1 - \delta_{a,b}) + ({\overline q} + \beta^{-1} \sum_{\omega \neq 0} {\overline D}_\omega e^{-{\rm i}\omega(\tau_1-\tau_2)}) \delta_{a,b}$ will be used together with $\psi^{(a)}_i(\tau) = \beta^{-1} \sum_\omega \psi_i^{(a)}(\omega) e^{-{\rm i}\omega \tau}$. The assumption of replica symmetry in $\omega=0$ terms is sufficient for the present purpose of showing a divergent 
conductivity because a breaking of replica symmetry would be accompanied by an independent parameter such as a coefficient of a quartic term 
on $Q^{ab}$. Further, only for the convenience of presentation, the on-site $\psi$-fluctuation (i.e., zero-dimensional case) with $c_g \to 0$ will be assumed because extending to higher dimensional and $H \neq 0$ cases can be trivially performed. Then, the variational equation $0= {\rm lim}_{n \to +0} n^{-1} \partial {\overline {Z^n}}/\partial Q^{(ab)}(\tau_1,\tau_2)$ is reexpressed by the following three equations:

\begin{equation}
\kappa^{-1}(\omega^2+r)-\kappa {\overline D}_\omega^2 + u({\overline q}+\beta^{-1}\sum_{\omega \neq 0} {\overline D}_\omega) - w_\psi G_d(\omega)=0,
\end{equation}
where $\omega \neq 0$, 
\begin{equation}
\kappa^{-1} r - \kappa \beta^2 ({\overline q}^2 - q^2) + u({\overline q}+\beta^{-1} \sum_{\omega \neq 0} {\overline D}_\omega) - w_\psi G_d(0)=0, 
\end{equation}
\begin{equation}
2 \kappa \beta^2 q ({\overline q}-q) + w_\psi G_{od}(0)=0. 
\end{equation}
Here, the SC fluctuation propagator $G^{(ab)}(\omega)= \beta^{-1}<(\psi_i^{(a)}(\omega))^* \psi_i^{(b)}(\omega)>$ is given by $G^{(ab)}(\omega) = \delta_{a,b}(1-\delta_{\omega,0}) G_d(\omega) + \delta_{\omega,0}(\delta_{a,b} G_d(0) + (1-\delta_{a,b})G_{od}(0))$, where
\begin{eqnarray}
G_d(\omega) &=& \frac{t}{r_\psi+c_\psi \omega^2 - w_\psi {\overline D}_\omega}, \\ \nonumber
\cr G_{od}(0)&=&\frac{w_\psi \beta q}{t} (d_\psi(\Delta q))^2, \\ \nonumber \cr G_d(0)&=& d_\psi(\Delta q) + G_{od}(0), \\ \nonumber
\cr d_\psi(\Delta q) &=& \frac{t}{r_\psi+w_\psi \Delta q}, 
\end{eqnarray}
and $\Delta q= \beta (q-{\overline q})$. Noting that, when $q > 0$, eq.(10) becomes 
\begin{equation}
\Delta q=\biggl(\frac{w_\psi}{t} \biggr)^2 \frac{t}{2 \kappa} (d_\psi(\Delta q))^2, 
\end{equation}
we easily find that the only physically meaningful solution of the PG order parameter is given together with eq.(12) by 
\begin{eqnarray}
{\overline q}= - \beta^{-1} \sum_{\omega \neq 0} {\overline D}_\omega + u^{-1} (\kappa (\Delta q)^2 + w_\psi d_\psi - \kappa^{-1} r), \\ \nonumber
\cr
{\overline D}_\omega = - \Delta q - \kappa^{-1} |\omega| \, \biggl(\frac{1+\kappa c_\psi w_\psi d_\psi^2/t}{1 + t \kappa^{-1} d_\psi^3 (w_\psi/t)^3} 
\biggr)^{1/2}.
\end{eqnarray} 
The above form of ${\overline D}_\omega$ is valid up to O($|\omega|$). In $w_\psi \to 0$ limit where $\psi$ and $Q$ fields are decoupled, the above solution reduces to the pure mean-field solution \cite{Phillips2,Read,Dalid} with $q > 0$ and $\psi=0$. The crucial point is that ${\overline D}_{\omega \to 0} \neq 0$ in the presence of the $\psi$-fluctuation. Situation is similar to the Ising spin-glass case in a magnetic field \cite{Read}. Actually, it will be recognized that, in eqs.(8) to (10), the fluctuation propagator $G^{(ab)}$ plays similar roles to  an external magnetic field in the spin-glass 
problem. Further, it will be clear that eq.(13) is also valid in higher dimensional case and $H \neq 0$ case if the expressions of $d_\psi^m$, where $m>0$ is an integer, are appropriately replaced. 

Let us turn to examining the conductivity $\sigma$ in terms of Kubo formula. Contributions to $\sigma$ arise from ${\cal F}_A$ and the $|D_i \psi_i|^2$ term in ${\cal F}_\psi$. Due to the $|\omega|$-dependence in ${\overline D}_\omega$, the latter contribution leads, as in Ref.\cite{Phillips}, to a {\it finite} contribution\cite{Phillips} to $\sigma$ in $T \to 0$ limit. The expression of ${\cal F}_A$ clearly implies the absence of {\it static} phase rigidity. However, it does {\it not} imply a finite conductivity because of $\beta {\overline q} \neq {\overline D}_{\omega \to 0} \neq 0$. The PG contribution to $\sigma$ arising from ${\cal F}_A$,  
\begin{eqnarray}
\sigma_{\rm PG}({\rm i}\omega) &=& \frac{4}{|\omega|} \lim_{n \to 0} \frac{1}{n} \sum_{a,b} \int_0^\beta d(\tau_1-\tau_3) e^{{\rm i}\omega(\tau_1-\tau_3)} \\ \nonumber 
\cr \times \biggl[ \int_0^\beta &d\tau_2& \delta(\tau_1-\tau_3) |Q^{(ab)}(\tau_1, \tau_2)|^2 - |Q^{(ab)}(\tau_1-\tau_3)|^2 \biggr],
\end{eqnarray}
is essetially the same as eq.(14) in Ref.\cite{Phillips2} and corresponds to the sum of $\sigma^{(1)}$ and $\sigma^{(2)}$ defined in Ref.\cite{Phillips2}.
The first term $\sigma^{(1)}$ results only from the static components $q$ and ${\overline q}$ and thus, should vanish consistently with the absence of static phase rigidity. Actually, it was verified \cite{Phillips2} through a delicate manipulation of an analytic continuation. We will focus rather on $\sigma^{(2)}$, which is given by
\begin{eqnarray}
\sigma^{(2)}(\! &{\rm i}& \!\omega) = \frac{4}{|\omega|} \biggl( \beta^{-1} \sum_{\omega_1} {\overline D}_{\omega_1} ({\overline D}_{\omega_1} - {\overline D}_{\omega_1+\omega}) \\ \nonumber 
\cr &+& \beta^{-1} (2 {\overline D}_\omega - {\overline D}_0) {\overline D}_0 - 2{\overline q} \, {\overline D}_{\omega} \biggr).
\end{eqnarray}
We are not interested in the first two terms under the $\omega_1$-summation which will lead to a finite contribution to $\sigma^{(2)}$. The last term in the bracket, $-2 {\overline q} \, {\overline D}_0$, is positive at low $T$ ($=\beta^{-1}$) 
limit and when $q > 0$ because ${\overline q} \simeq q -$ O($T$). Since, according to eq.(13), the second two terms in the bracket also give only a O($T$) correction to the above term, a divergent conductivity in low $T$ limit results from eq.(15), as in the Meissner phase, as far as the PG order is present (i.e., $q > 0$) as a consequence of a nonzero $J_0 \propto w_\psi$. 

Finally, we briefly show that this divergence of conductivity due to the PG order is preceded by that due to another SC glass ordering induced by the fluctuation of PG order parameter $Q^{ab}$. This SC glass corresponds to the vortex-glass \cite{Fisher} studied so far for nongranular materials in $H \neq 0$, and we closely follow below the treatment in Ref.\cite{RI3}. Let us relax the infinite-range approximation for the disorder part ($\propto J^2$) of the action to incorporate a gradient term $(\nabla Q^{ab})^2$. Using results in Ref.\cite{Read} in the $q=0$ case and integrating over the $Q^{ab}$-fluctuation in the Gaussian approximation, one finds that the nonlinear terms in $\psi$ of ${\cal F}_\psi$ take the form \cite{Lubensky} 

\begin{eqnarray}
\Delta {\cal F}_\psi &=& \frac{\beta^{-3} w_\psi^2}{2 t} \sum_a \prod_{i=1}^4 \sum_{\omega_i} \int_{\bf k} \delta_{\omega_1+\omega_2, \omega_3+\omega_4} \\ \nonumber \cr
\times ( u_\psi &-& G^c({\bf k}, \omega_1, \omega_3, \omega_2) ) \, \rho_{\omega_1, \omega_3}^{(aa)}({\bf k}) \, \rho_{\omega_2, \omega_4}^{(aa)}(-{\bf k}) \\ \nonumber \cr
&-& \frac{\beta^{-2} w_\psi^2}{2 t} \sum_{a,b} \sum_{\omega_1, \omega_2} \int_{\bf k} G({\bf k}, \omega_1, \omega_2) \, \rho_{\omega_1, \omega_2}^{(ab)}({\bf k}) \, \rho_{\omega_2, \omega_1}^{(ba)}(-{\bf k}), 
\end{eqnarray}
where $u_\psi > 0$, and \cite{Read} 
\begin{eqnarray}
G^c({\bf k}, \omega_1, \omega_3, \omega_2) &=& - \frac{u}{t} G({\bf k}, \omega_1, \omega_3) G({\bf k}, \omega_2, \omega_4) \\ \nonumber \cr
\times \biggl( 1 &+& u \beta^{-1} \sum_\Omega G({\bf k}, \Omega+\omega_1, \Omega+\omega_3) \biggr)^{-1}, \\ \nonumber \cr 
G({\bf k}, \omega_1, \omega_2) &=& \frac{t}{k^2 + \sqrt{\omega_1^2 + {\tilde r}} + \sqrt{\omega_2^2 + {\tilde r}}}, 
\end{eqnarray}
where $\rho_{\omega_1, \omega_2}({\bf k})$ is the Fourier transform of $(\psi_i(\omega_1))^* \, \psi_i(\omega_2)$, and the length scale was properly 
normalized. The $-G^c$ term in eq.(16), due to its tendency enhancing with decreasing ${\tilde r}$, dominates over the original $u_\psi$ term and suggests a great {\it enhancement} \cite{RI2} of quantum SC fluctuation due to the granular structure. The onset of PG order is signaled, within the Gaussian fluctuation, by ${\tilde r} \to +0$, while the SC glass ordering is signaled by a 
divergence of the glass susceptibility \cite{Fisher} 
\begin{equation}
\chi_{\rm sg} =N^{-2} \sum_{i,j} {\overline {|\langle \psi_i(\omega=0) (\psi_j(\omega=0))^* \rangle|^2}}.
\end{equation}
If expressing $\chi_{\rm sg}$ by ladder diagrams like $\chi_{\rm sg} = 1 + \sum_{n=1} I^n$, its irreducible vertex $I$ is proportional to the vortex-pinning energy term, i.e., the minus second term of eq.(16), with vertex corrections due to the $\psi$-interaction term (the first term of eq.(16)). Let us restrict ourselves to the lowest Landau level (LLL) of the $\psi$-fluctuation and assume the presence of PG order in low $T$ limit (i.e., ${\tilde r}(T \to 0) \to +0$). Since the $\psi$-fluctuation in LLL is noncritical at low $T$ limit, a ${\tilde r}$-dependence of $I$ primarily determines $\chi_{\rm sg}$. If the interaction vertex corrections are neglected, we find that $I$ in LLL is divergent upon cooling like ${\tilde r}^{-1/4}$, suggesting that the SC glass ordering occurring prior to the PG ordering induces a divergence of conductivity. Hence, it is important to verify whether this divergence is suppressed or not by the interaction vertex corrections. If calculating this vertex correction in a consistent way with the Hartree approximation \cite{RI3,David}, the strength $\propto r^{-1/4}$ of the resulting renormalized vortex-pinning energy is modified in the manner 
\begin{equation}
{\tilde r}^{-1/4} \to {\tilde r}^{-1/4} \biggl( 1 - \frac{v_{q_y} G(q_y {\hat y}, 0,0) \beta^{-1} \sum_\omega G(q_y {\hat y}, \omega, \omega)}{1 + v_{q_y} \beta^{-1} \sum_\omega (G(q_y {\hat y}, \omega, \omega))^2} \biggr)^2,
\end{equation}
where $v_{q_y} = e^{-q_y^2/2}$, and $q_y$ is an external wave number carried by the $\psi$-fluctuation propagator. Further, the factor in the bracket of eq.(17) was neglected because it plays no roles in ${\tilde r} \to 0$ limit, and the positive coefficients $w_\psi$, $u$, and $t$ were set to be unity because their detailed values are unimportant here. Focusing on the $T \to 0$ limit, one easily verifies that the vertex correction merely brings an enhancement factor $\sim ({\rm ln}[{\rm min}(q_y^2, \sqrt{{\tilde r}})])^2$. Thus, the result obtained without the interaction vertex correction is essentially unaffected, and, even in $H > {\overline H}_{c2}(0)$, the SC transition (vanishing of resistivity) in nonzero fields is expected to be driven by the vortex-glass ordering induced by an effective vortex-pinning effect. 

This result, independent of the dimensionality of systems, is apparently similar to but different from the main conclusion in Ref.\cite{GaL}. In Ref.\cite{GaL}, an upper limit of the SC transition field at $T=0$ is obtained above ${\overline H}_{c2}(0)$ within the range where a random XY model similar to eq.(1) is applicable. On the other hand, the critical PG fluctuation inducing the SC ordering in the present model exists in any $H$ as far as the granular model is valid and the SC fluctuation is present. Namely, an upper limit of $T=0$ SC transition field in the present case is roughly given by the field at which the model eq.(1) breaks down. 


%

\end{document}